\documentclass[9pt,twocolumn,twoside]{osajnl2}
\journal{ao} 

\setboolean{shortarticle}{false} 

\usepackage{amsmath}
\usepackage{graphicx}
\usepackage[colorinlistoftodos]{todonotes}

\title{Polarization Independent Atomic Prism  filter for removing Amplified Spontaneous Emission}
\author[1,*,+]{Raphael David Cohen}
\author[1,2,*]{Christopher A. Mullarkey}
\author[1,2]{John C. Howell}
\author[1]{Nadav Katz}
\affil[1]{Racah Institute of Physics, The Hebrew University of Jerusalem, Jerusalem, Israel, 91904}
\affil[2]{Department of Physics and Astronomy, University of Rochester, Rochester, New York, USA, 14627}
\affil[*]{equal contributions}
\affil[+]{Corresponding author: raphael.cohen3@mail.huji.ac.il}
\ociscodes{(130.7408)  Wavelength filtering devices, (020.1335)   Atom optics, (140.2020) Diode lasers, (230.5480)  Prisms, (350.2450)   Filters, absorption, (270.5585) Quantum information and processing.}

\begin{abstract}
We  create an optical frequency, polarization independent, narrow band-pass filter  of  1.3 GHz (3 dB bandwidth), using the steep dispersion near the Rubidium D$_1$ atomic transitions within a prism-shaped vapor cell. This enables us to clean the amplified spontaneous emission from a laser by more than 3 orders of magnitude. Such a filter could find uses in fields such as quantum information processing and Raman spectroscopy.

The published article can be found in Applied Optics Vol 57 Issue 16

\url{https://www.osapublishing.org/ao/abstract.cfm?uri=ao-57-16-4472}

\end{abstract}

\setboolean{displaycopyright}{true}

\begin{document}
\maketitle
\section{Introduction}
A laser amplifies  stimulated emission using a gain medium and a cavity. However when the gain media is broadband one inevitable  side effect will be the amplification of spontaneous emission, (ASE) \cite{amnon1989quantum}. This ASE appears as a low power broadband pedestal to the relatively high power narrow bandwidth lasing signal, all in the same spatial mode. The precise amount of ASE will depend on the gain media, cavity and operating conditions of the laser. For some applications, the ASE is a desirable  source of incoherent light in a well defined spatial mode\cite{737626}, while for most applications this ASE will appear as a small amount of noise.
    
    However, there are experiments, such as in Raman spectroscopy and in quantum information processing, where this background (typically rated at $<-30$ dB) can still significantly affect the SNR and special methods must be employed  to reduce it further.  For example, consider a situation like the DLCZ protocol where a control pulse produces a quantum signal \cite{Duan_Nature,RevModPhys.83.33}, in a very close spatial and frequency mode which must be filtered. In situations like these even with 60dB suppression of ASE there will be 1pW of ASE for every mW of laser, producing  $\mathcal{O}(10^{6})$ photons every ms,  which can be a very large background if your signal is one photon!  
 For these highly sensitive applications two main technologies are used to increase SNR, specialized Bragg gratings \cite{Moser2009,OndaxCleanLine2}, and Fabry-Perot etalons \cite{Bashkansky:12}. Specialized gratings can reach a transmission width of around 50 GHz, whereas etalons can have widths of 100 MHz, but with less suppression \cite{NoteInhomFilter}.
 
 In this paper we take one of the most well known optical elements, a  prism \citep{newton1687philosophiae}, and  replace the glass with a relatively dense  $10^{14} $ cm$^{-3}$ atomic vapor. At frequencies close to atomic transitions a large change in refractive index occurs $\mathcal{O}(0.001)$ over a very small frequency $\mathcal{O}(\text{GHz})$ producing a dispersive power many orders of magnitude larger than glass. With this effect we separate laser light from ASE even up to few hundred MHz. In Section  \ref{sec:Theory} we outline the basic theory, then in Sections \ref{sec:Method} \& \ref{sec:Res} we describe our preliminary implementation, followed by a discussion in Section \ref{sec:Discuss} and concluding remarks in Section \ref{sec:Conc}.

\section{Theory}\label{sec:Theory}

The refractive index of an atomic ensemble is directly related to the susceptibility, (see Eq. \ref{eq:X}). This susceptibility is modeled by the convolution of the Lorentzian atomic transitions with the Doppler broadening, known as the Voigt profile, together with a strength mainly proportional to the atomic density, 	Eq. \ref{eq:n2}, based upon \cite{siegman1986lasers}. Rubidium has a highly temperature dependent vapor pressure \cite{Steck85_2}. Heating to a high temperatures ($\approx150 ^\circ$C) has been demonstrated to create a vapor with a dispersive power many orders of magnitude larger than standard materials \cite{PhysRevA.86.023826,PhysRevA.87.043815}. 

\begin{eqnarray}\label{eq:X}
n^2(\omega) &=& 1+\chi(\omega)\\
\nonumber\label{eq:n2}
 &=& 1+N(T)D_N\int\frac{e^2}{2\omega_0m\varepsilon(\omega-\omega'+i\gamma/2)}\\
 & &\hspace{1.5cm}\times\exp\left[-4ln2\left(\frac{\omega'-\omega_0}{\Delta\omega_d}\right)^2\right]d\omega',
\end{eqnarray}

where the temperature dependent number density is $N(T)=P(T)/k_BT$, where the pressure, $P(T)$, is given in  \cite{VapourPressureEqns,Steck85_2}, the Doppler width is $\Delta\omega_d= \sqrt{\frac{(8ln2)k_BT}{mc^2}}\omega_0$, and the normalization constant is $D_N=\sqrt{\frac{4ln2}{\pi\Delta\omega^2_d}}$ for a transition frequency $\omega_0$.
Using this as the dispersive medium in a prism, light which is close in frequency to an atomic transition,  will undergo refraction. In contrast, light that is detuned  will be unaffected and pass straight through, (Fig. \ref{fig:layout}). Focusing this deflection $\delta(\omega)$ by a lens of focal length $f$ will give a shift $s$. Using the small angle approximation and Snell's law for a right angled prism yields

\begin{eqnarray}\label{eq_shift}
\nonumber
s &=& f\delta(\omega) \\
\nonumber
&=&f\Big[\sin^{-1}\left(\sin\theta_1\cos\phi+\sin\phi\sqrt{n^2(\omega)-\sin^2\theta_1}\right)\\
& &\hspace{4cm}-\theta_1-\phi\Big].
\end{eqnarray}

A spatial filter, such as a pinhole, placed at the location of the shift will now act as frequency filter, selecting  the frequency determined by the pinhole. Light at this frequency will be transmitted, while other frequencies (the ASE) will be blocked.  
To find the best operating parameters we define a sensitivity parameter $\eta$ that is the ratio of the shift and beam width $\sigma(L)$ at distance $L$ from the focusing lens. Assuming a collimated Gaussian beam we arrive at 

\begin{eqnarray}\label{eq:sensitivity}
\nonumber
\eta &=& s/\sigma(L)\\
&=&\frac{L\delta(\omega)}{\sigma_0\sqrt[]{\left(1-\frac{L}{f}\right)^2+\left(\frac{L}{z_R}\right)^2}}.
\end{eqnarray}

This will have a maximum at $L=f$
\begin{eqnarray}\label{eq:sensitivity_max}
\eta_{max} = \frac{\delta(\omega)z_R}{\sigma_0}=\frac{\delta(\omega)\pi\sigma_0}{2\lambda_0}.
\end{eqnarray}

Interestingly we see that the maximum sensitivity is at the focal length  and not at the slightly closer position of the new beam waist. Also we see that for a fixed prism angle, the only way to increase the sensitivity is to work at higher temperatures (greater $\delta(\omega)$) or a larger beam diameter. 

\section{Method}\label{sec:Method}

To practically demonstrate this effect, we work with the alkali metal Rubidium in a natural abundance vapor cell at the D$_1$ transitions (795 nm). For the prism, we use a cylindrical cell (diameter=2.54 cm) cell with angled faces $(45^{\circ})$. To obtain sufficient atomic density, we heat the cell by placing it in an aluminum box and wrapping the bow with about 14 winds of Nikrothal 60. We then apply up to 54 W of power to heat the device to around $150^{\circ}C$. Entry and exit openings for the laser are left in the box, as well as for a "cold" spot at a distance of a few centimeters from the prism regime \cite{NoteColdspot}. This creates a location for a point of condensation away from the prism when the Rubidium cools, and where we measure the temperature of the cell (using a Thorlabs TH10K thermistor).
\begin{figure}
\centering
\includegraphics[width=0.4\textwidth]{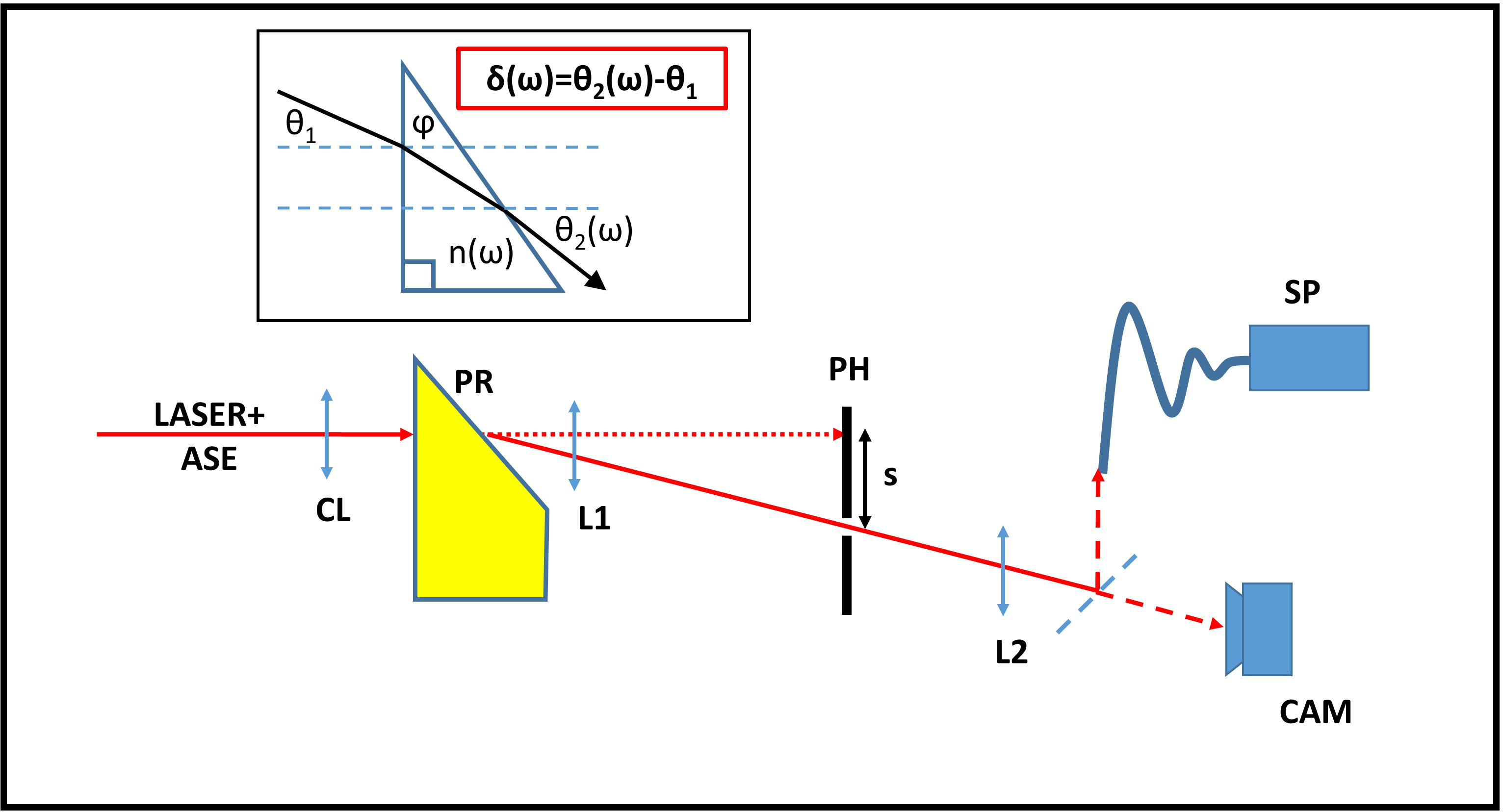}
\caption{Schematic of experimental set-up :  CL-cylindrical lens, PR-prism, at end of heated Rubidium cell, L1-focusing lens, PH-pinhole, L2-imaging lens, CAM- Camera, SP- spectrometer (switched with photodiode), s- shift of laser. The inset shows the shift of light by a prism.}\label{fig:layout}
\end{figure}
We work with two TOPTICA lasers. The first is a DFB, with which we work with non-standard parameters (current slightly above lasing threshold) to increase the ASE ratio (from $\approx$-30 dB to $\approx$-10 dB), without doing this we would not  have sufficient dynamic range in our equipment for the measurements. The second laser is a DLPro (with an ASE ratio of $\approx$-60 dB), which effectively acts as an ASE free source (for our sensitivity). 
The laser enters through the curved side of the cell, and exits through the flat angled face. To compensate for the lensing effect of the curved cell wall, we place a cylindrical lens before the prism. The laser light is coupled into a single mode fiber. After exiting the fiber we collimate the light (diameter = 2.1mm) before it passes through a cylindrical lens  upon entering the prism. After exiting the light is focused. An imaging lens in conjunction  with a mirror on a flipper enables us to image this plane onto a camera (Point Grey Firefly MV) or to  observe the signal in a spectrometer (Ocean Optics HR2000+) or photodiode (Thorlabs PDA36A), see Fig. \ref{fig:layout}.

\section{Results}\label{sec:Res}

\begin{figure}
\centering
\includegraphics[width=0.4\textwidth]{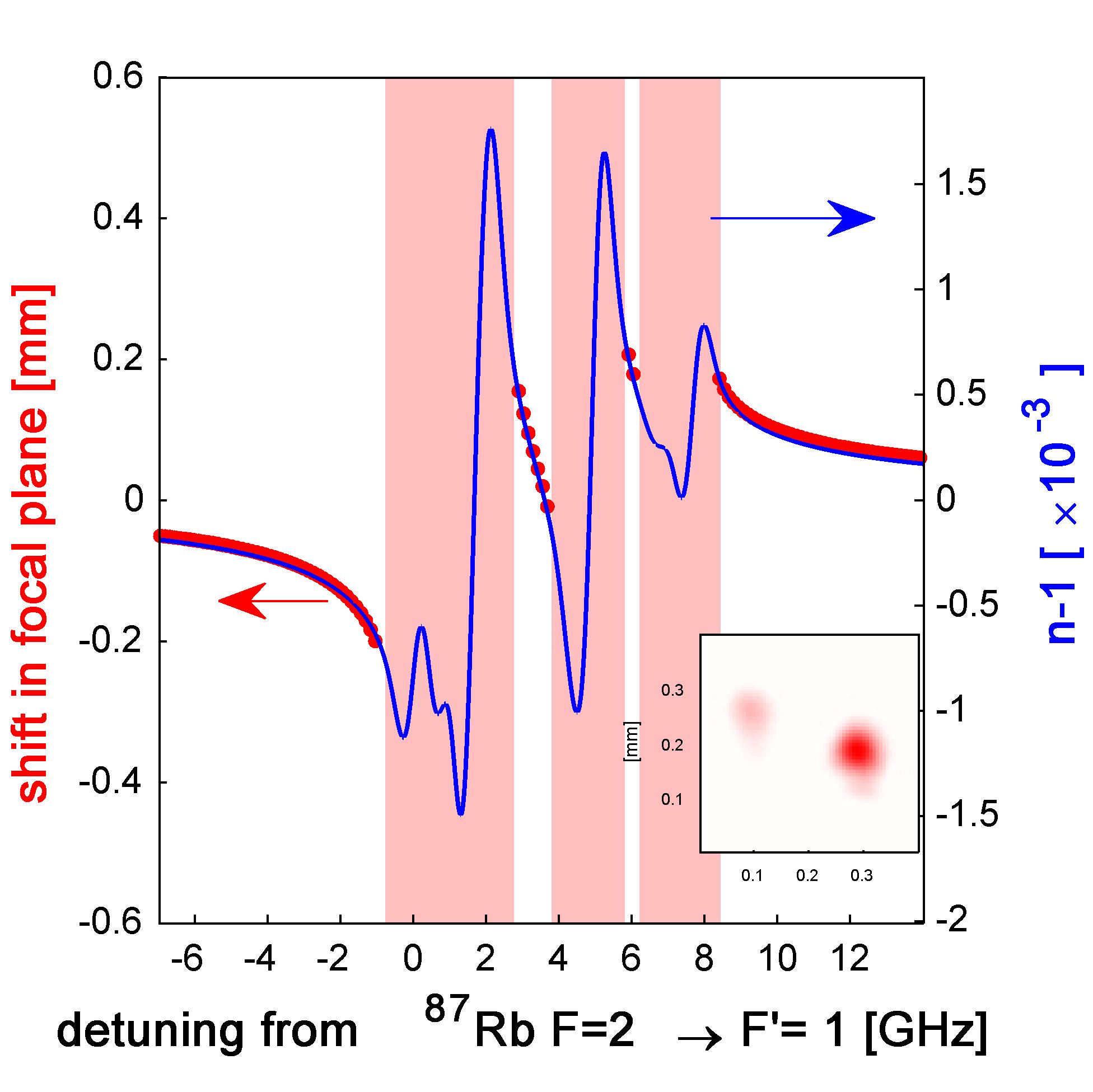}
\caption{The laser light is shifted from the ASE. The main figure shows the predicted refractive index change  (solid blue), and  measured shift (red dot). The shaded regions are of absorption greater than 50 \% . Calculations are for $143 ^{\circ}$C and f=300 mm, $\phi$=$45^{\circ}$,  $\theta$=$0^{\circ}$. The inset  shows the shift of laser light close to resonance (weaker) and from ASE (stronger). For filtering a pinhole is placed at the location of the laser light.}
\label{fig:shift_pic}
\end{figure}

First, we demonstrate the ability of the prism to cause  a frequency-dependent deflection. This deflection is observed to be independent (to within 5\%) upon rotation  of a $\lambda$/2 waveplate. Polarization independence is entirely consistent with the theory, Sec. \ref{sec:Theory}, which, due to the relatively large detuning and lack of magnetic fields, is insensitive to the subtleties of the Rubidium hyperfine structure.
The  inset in Fig. \ref{fig:shift_pic} shows the laser light being separated from the ASE by approximately 8$\sigma$. Despite the inherent 10dB suppression of the ASE, when all frequencies are considered  there is more power in the ASE than at the lasing frequency (because we are operating with the DFB laser just above threshold). The small shift in the vertical direction  arises from our optical set-up using a curved cell wall. As we heat the prism small movements occur which misalign our compensation for the curved cell wall, that we are unable realign due to the high temperatures. Using our displacement model based upon Eq. \ref{eq:n2} \& \ref{eq_shift} while accounting for the eight different Rb $D_1$ transitions, allows us to model the shift of the laser to a high degree of accuracy, Fig. \ref{fig:shift_pic}. Optimal fitting (blue) is achieved by small adjustments to the temperature of the model ($\mathcal{O}$(10$^\circ$C), as the measured temperature appears to be higher than the actual temperature of the vapor \cite{NoteColdspot}. Frequency dependent deflections of the laser light were observed to be independent of laser power (<1$\mu$W $\rightarrow$ 30mW). Although one may expect that at higher laser intensity there will be optical pumping that effectively reduces the atomic density, the large detuning means that there is  corresponding reduction in the scattering rate 
\begin{eqnarray}\label{eq:scattering rate}
\nonumber
R=\frac{\Gamma}{2}\frac{I/I_{sat}}{1+4(\Delta/\Gamma)^2+(I/I_{sat})},
\end{eqnarray}
where $\Gamma$ is the spontaneous emission rate, I is the laser intensity, $\Delta$ is the detuning, and $I_{sat}$ is the saturation intensity \cite{Steck85_2}.  A simple model considering the transit time of an atom transversing the beam shows that for our experimental parameters we could use 200mW of  power without  a significant  (<5\%) change in deflection. 

\begin{figure}
\centering
\includegraphics[width=0.5\textwidth]{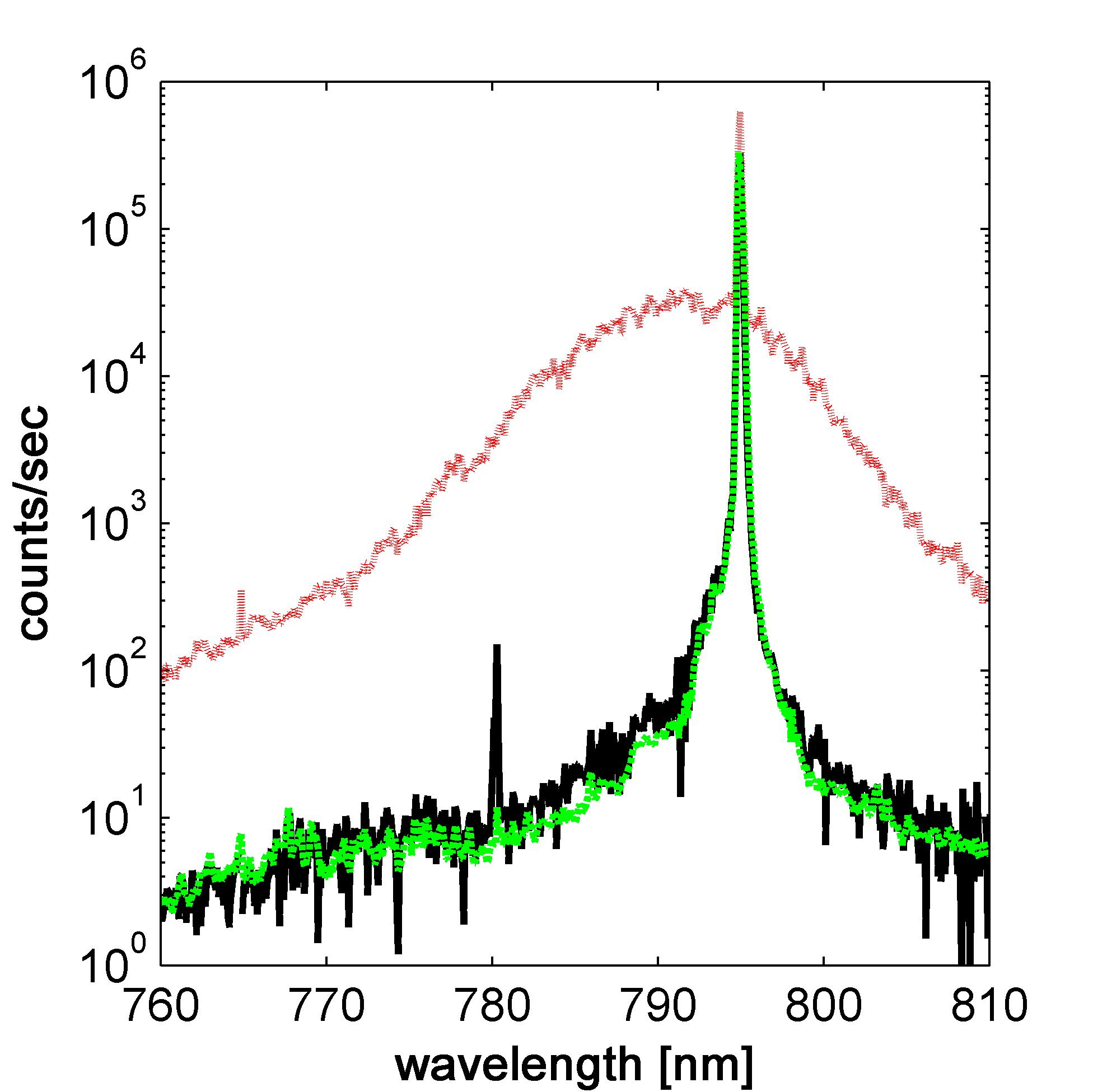}
\caption{\label{fig:ASE_sup}ASE suppression. The red (dotted) line shows the profile of the laser without the prism and pinhole. The black (solid) line shows the spectrum of the light with the pinhole. The green (dashed)  line shows the spectrum of a laser with  60 dB ASE suppression}
\end{figure}

Next we place a 100 $\mu$m diameter pinhole at the location of the laser light in the focal plane, transmitting the  laser and blocking the ASE. The transmission efficiency is $\approx$40\%, (this is not intrinsic but related to mismatching between the pinhole and the beam shape). Increasing the pinhole radius will transmit more laser light at the expense of letting in more ASE. For a Gaussian beam, the transmitted light will be 
\begin{eqnarray}\label{eq:transmitted light}
\nonumber
T_{laser}=2\pi\int_0^a{I_{laser}re^{-2r^2/\omega'^2}dr},
\end{eqnarray}
while the ASE is deflected by $s$, Eq. \ref{eq_shift}, and the transmitted component will be given by,
\begin{eqnarray}\label{eq:transmitted light2}
\nonumber
T_{ASE}=\int_{s-a} ^{s+a}\int_{-y'}^{y'}{I_{ASE}e^{-2(x^2+y^2)/\omega'^2}dxdy},
\end{eqnarray}
where $\omega'$ is the new waist of the focused beam, a is the radius of the pinhole and the parameter for the integration over y is  $y'=\sqrt{a^2-(x-s)^2}$. However, in practice real laser beam shapes need to be considered in the transmission/filtering tradeoff, which in our experiment was also influenced by the optics (our curved cell wall and compensation). 

We now compare the spectrum of the light  with and without the pinhole,  Fig. \ref{fig:ASE_sup}. The red (dotted) line shows the spectrum without any pinhole, that is with ASE, while the  black (solid) line shows the spectrum with the pinhole, that is the cleaned laser light. We believe that the spread in the cleaned laser light is a limitation of the spectrometer's point spread function and not from ASE. This is seen on the DLPro laser which is effectively clean from ASE on this scale. The green (dashed) line shows that the DLPro spectrum, after normalization, which essentially follows the background of the cleaned laser light \cite{NotePSF}.  This places the background at a few counts/sec and an ASE suppression of greater than 30 dB. It is interesting to note the small signal at 780 nm which comes from a small amount of ASE at the frequency of Rubidium D$_2$ transitions which is deflected into the pinhole. The  focusing lens (L2 in Fig. \ref{fig:layout}) collects the spontaneous emission according to the solid angle from the cell, of this light only a small fraction comparable to the ratio of the pinhole to lens area will enter the measurement. We placed the  focusing lens at a distance of approximately 30 cm to reduce this effect by >60 dB.

\begin{figure}
\centering
\includegraphics[width=0.5\textwidth]{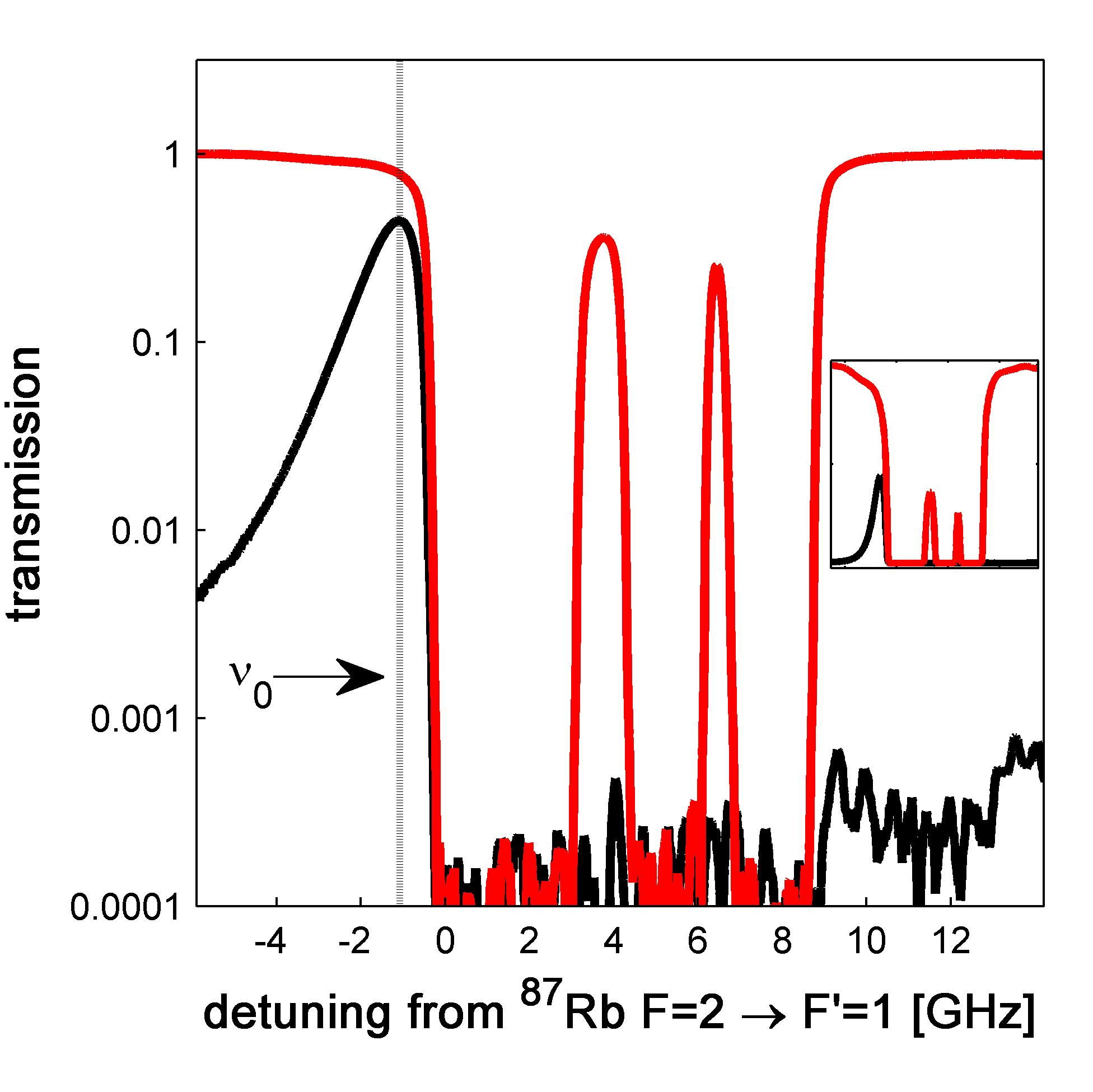}
\caption{The main figure shows a logarithmic plot of the pinhole transmission as a function of frequency. The red line shows the transmission without the pinhole, while the black line shows the transmission with the pinhole. The inset shows the non-log plot of the same data. }\label{fig:pinhole}
\end{figure}

Finally, in order to measure the width of of our bandpass filter,  we replace the spectrometer with a photodiode. We use the clean DLPro laser, scanning the frequency with corresponding spatial shift and measuring the transmission through the pinhole. The results,  Fig. \ref{fig:pinhole},  clearly demonstrate the narrowness (1.3 GHz, 3 dB) of the filter, with a >30 dB suppression far from the transmission frequency, $\nu_0$. As expected the filter is asymmetric. We have chosen $\nu_0$ to be  red detuned from the D$_1$ lines, this means that at frequencies greater than $\nu_0$ there is a rapid falloff due to the resonant absorption of the Rubidium. Even at frequencies greater than the Rubidium  D$_1$ lines there is still a shift of the light in the direction away from the pinhole which aids the suppression. When the frequency is decreased from $\nu_0$ the opposite is true, with the light shifted towards the pinhole and hence causing a slower falloff in the suppression.

\section{Discussion}\label{sec:Discuss}
Our preliminary experiments clearly demonstrate the ability of the prism to reduce ASE by 3 orders of magnitude, Fig. \ref{fig:ASE_sup} \& \ref{fig:pinhole}, which approaches the limit of the dynamic range of our equipment. 

\begin{figure}
\centering
\includegraphics[width=0.4\textwidth]{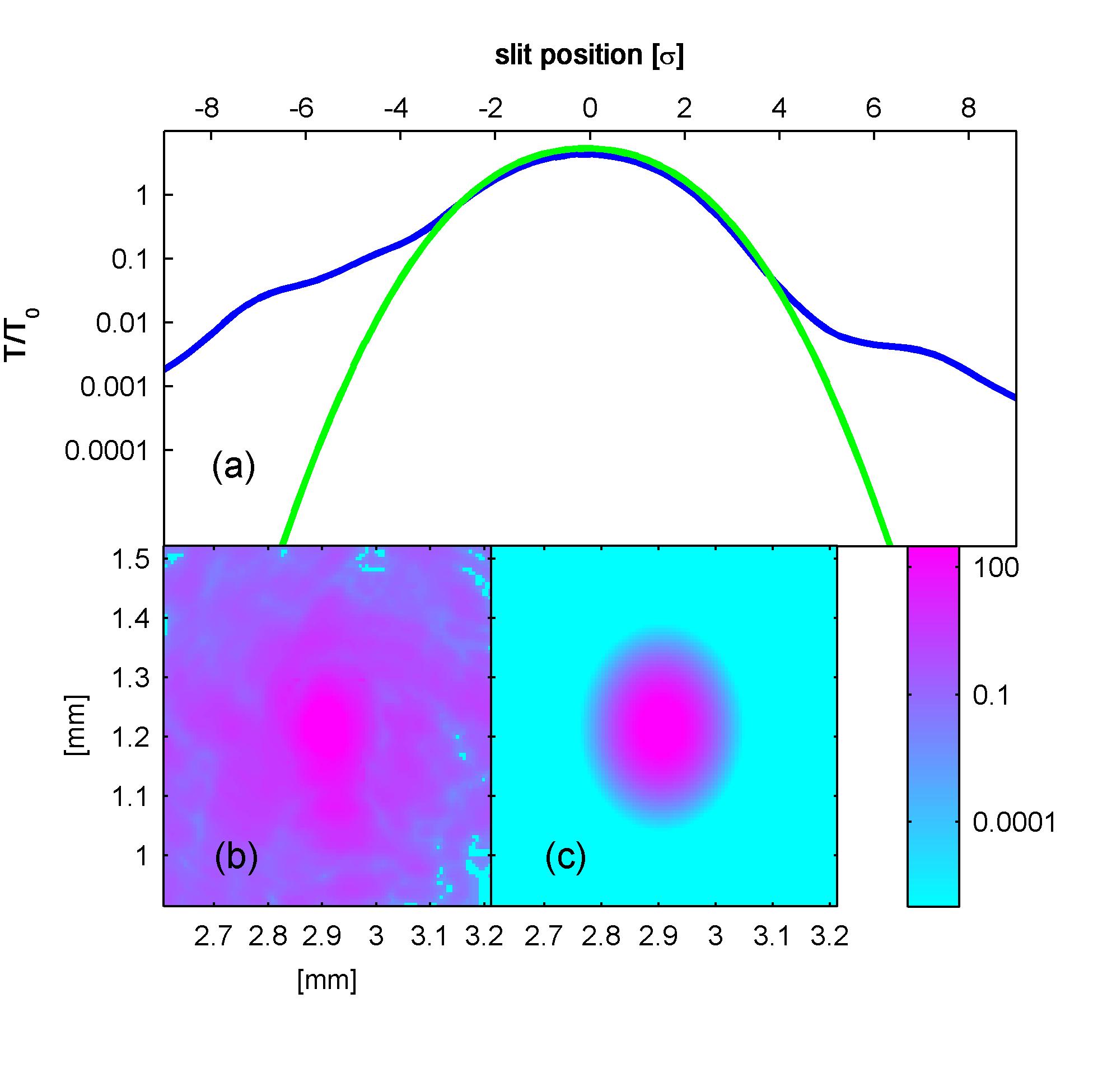}
\caption{The main figure (a) shows the expected fraction of beam intensity that would pass through a pinhole, as the location pinhole moves in the x direction, away from the center. The blue curve represents the expected transmission for the actual beam, whereas the green line shows expected transmission for a 2D Gaussian fit to the measured profile. A log plot of the measured laser image can be seen in the bottom left figure (b) (corresponding to the blue line in (a)).  The log plot of the 2D Gaussian fit to this image can be seen in the bottom right figure (c) (corresponding to the green line in (a)).} \label{fig:beam_comp}
\end{figure}\label{fig:ph_theoretical}

Upon investigation we also found that  the suppression was also limited by the laser beam profile. A truly Gaussian beam will have a much sharper fall off in intensity than we observed, see Fig. \ref{fig:ph_theoretical}. We should observe the same ASE suppression (30dB) for a shift corresponding to 3-4$\sigma$ however we required shift of about 8$\sigma$. This is mainly explained by the spatial mode quality. Even though our spatial mode is a very good approximation to a Gaussian beam, at radial displacements greater than a  few $\sigma$ there is a much slower fall-off in the intensity than expected, see the insets in Fig. \ref{fig:ph_theoretical}.  This limits the effectiveness of the spatial filter, requiring a bigger shift to separate the shifted spatial mode from the original mode. 

Even with our non-ideal beam our results compare very favorably  to the other options for narrow pass optical filters. Our filter is 50$\times$ narrower than the best ASE filters known to us, \citep{OndaxCleanLine2} while the suppression ratio is hard to compare with the available data. 
Extremely narrow  filters with kHz bandwidth have been made, however these require coherent media such as EIT \cite{PhysRevA.81.063824,BI20164022}. High finesse etalons can reach 100's MHz, but they do not come close in their suppression ratio, especially considering  the repetitive nature of their pass band. Our filter, as a forward propagating spatial block also prevents  the issues associated with the reflections and distortions of high finesse cavities, especially at higher powers.
The prism  also allows for many different spatial modes of the light, but where a spatial extension in the direction of the shift will decrease the filter efficiency. 

As with almost all Atomic Resonance Filters (ARF) \cite{Gelbwachs1988,Yeh:82}, there is an obvious limitation in that we have a very small region of tunability. The prism works most effectively at both the blue and red sides of the Doppler broadened absorption, although by appropriate positioning of the pinhole it may be used at any frequency where there is sufficient deflection for the required application \cite{NoteTunability}. It should be noted that many applications allow for a compatible  frequency to be chosen \citep{Zentile:15}. Also in the field of  quantum information processing, a frequency close to an atomic resonance is desirable \cite{Zielinska:14}. One often has the ability to choose an appropriate isotope for the filter corresponding to the atoms being used material e.g. $^{87}$Rb for filtering a wavelength on the $^{85}$Rb transitions \cite{Bashkansky:12}. Other small frequency changes could be accomplished by devices such as an AOM. It should be noted that for a  quantum information experiment the small signal at a very far detuned transition (in our experiment the small signal at the Rubdium D$_2$ (780nm) transition frequency, see Fig. \ref{fig:ASE_sup}) can be easily filtered with a standard large bandwidth filter. 
Our atomic prism filter compares favorably with other ARFs. It is simple to implement, has a single pass band, is polarization independent, can work over a wide range of powers, and does not require a large atomic cell or magnetic field. Aside from the narrow 3 dB bandwidth of 1.3 GHz, Sec. \ref{sec:Res}, other measures of performance  such as equivalent noise bandwidth, ENBW$\approx1.5 \text{ GHz}$ and figure of merit FOM $\approx0.3 \text{ GHz}^{-1}$ ( $\text{ENBW}={\int T(\nu) d\nu}/{T(\nu_s)}, \text{ FOM}={\text{T}_{max}}/{\text{ENBW }} $\cite{Zentile:15}) , compare competitively with the state of the art for Faraday rotation filters \cite{Zielinska:12}. The potential for very high performance exists across a broad range of atoms and their resonances.

\section{Conclusion}\label{sec:Conc}
We have presented a novel method to create a high suppression (>30  dB) ultra narrow (1.3 GHz) optical frequency pass filter. We have demonstrated it to suppress ASE. Theoretically the transmission can approach unity, while the suppression can be many orders of magnitude. Such a filter working near atomic resonances could find useful applications in quantum information processing and other areas.   

\section{Acknowledgements}\label{sec:Ack}

Funding: ISF Grant No. 1567/12, Hebrew University Visiting Research Fellowship.

R.D.C. would like to thank Dr A. J. Cohen for his assistance with typesetting.

\newpage
\bibliographystyle{plain}
\bibliography{Prism.bib}

\begin{thebibliography}{10}
\bibitem{NoteInhomFilter}
A relatively intense narrow bandwidth laser can be passed through an
  inhomogeneously broadened media to remove ASE close to the lasing frequency.
  However such filtering will only remove the ASE very close to resonance as
  opposed to broadband schemes considered in this article.
\bibitem{NoteColdspot}
A hole in the metal casing enclosing the Rubidium cell creates a cold spot so
  that condensation occurs away from optically active surfaces. This cold spot
  is also used for temperature measurements, however as the heating wires are
  on the outside of the casing the Rubidium vapor ends up at a lower
  temperature than the measured temperature.
\bibitem{NoteTunability}
The specific atomic hyperfine atomic structure will determine if there is a
  preference for blue or red detuning. Greater deflections, and hence filtering
  can be accomplished by optimizing the prism angle, combing prisms and
  increasing the atomic density, although this may come at the expense of
  reduced transmission.
\end{thebibliography}

\end{document}